\documentclass{ws-ijmpcs}

\def\trimmarks{}

\usepackage{afterpage}

\newcommand{\gev}{\operatorname{GeV}}

\newcommand{\vek}[1] {\boldsymbol{#1}}

\newcommand{\y}{\vek{y}}

\newcommand{\ms}{\mskip 1.5mu}

\allowdisplaybreaks[3]

\begin{document}

\markboth{T. Kasemets}
{Scale evolution of double parton correlations}

%
\catchline{}{}{}{}{}
%

\title{Scale evolution of double parton correlations}

\author{Tomas Kasemets}

\address{Nikhef and Department of Physics and Astronomy, VU
  University Amsterdam, De Boelelaan 1081, 1081 HV Amsterdam, The
  Netherlands\\
  kasemets@nikhef.nl}

\maketitle

\begin{abstract}
We review the effect of scale evolution on a number of different correlations in double parton scattering (DPS). The strength of the correlations generally decreases with the scale but at a rate which greatly varies between different types. Through studies of the evolution, an understanding of which correlations can be of experimental relevance in different processes and kinematical regions is obtained. 
\keywords{double parton scattering; evolution; correlations.}
\end{abstract}

\section{Introduction}
An increasingly relevant aspect of proton-proton collisions at high energies is double parton scattering (DPS), where two partons from each proton interact in two separate hard subprocesses. DPS contributes to many final states of interest at the LHC. They constitute relevant backgrounds to precise Higgs boson coupling measurements and searchers for physics beyond the Standard Model.  Our knowledge of DPS is fragmentary, and improvements are needed at both conceptual and quantitative level (see for example \cite{Diehl:2014doa}).

Schematically the DPS cross section can be expressed as
\begin{align}\label{eq:cross}
\frac{d\sigma}{\prod_{i=1}^2 dx_i d\bar{x}_i }\Bigg|_{DPS} & = \frac{1}{C} \hat{\sigma}_1\hat{\sigma}_2 \int d^2\vek{y}\ F(x_1,x_2,\vek{y}) \bar{F}(\bar{x}_1,\bar{x}_2,\vek{y}),
\end{align}
where $\hat{\sigma}_i$ represents hard subprocess $i$, $C$ is a combinatorial factor equal to two (one) if the partonic subprocesses are (not) identical and $F$ ($\bar{F}$) labels the double parton distribution of the proton with momentum $p$ ($\bar{p}$). The DPDs depend on the longitudinal momentum fractions of the two partons $x_i$ ($\bar{x}_i$) and the distance between them $\vek{y}$. Implicit in this expression are the labels for the different flavors, colors, fermion numbers and spins of the four partons. This quantum-number structure is significantly more complicated in DPS compared to the case with only one hard interaction, because of the possibility of interference between the two hard interactions and correlations between the two partons inside each proton. 

The correlations can be of kinematical type (between $x_i$'s and $\y$), or between the quantum numbers of the two partons. While the kinematical type affects dependence of the DPDs on the kinematical variables, the quantum-number correlations lead to a large number of different DPDs. The DPDs depend on long distance, non-perturbative physics and can thus not be calculated in perturbative QCD.

Including all the correlations and their DPDs in phenomenological calculations is cumbersome, and extracting all of them experimentally is unfeasible. An effective way of reducing the number of DPDs of experimental relevance is studying the scale evolution of the DPDs and the correlations they describe. For example, it has been demonstrated for quark and antiquark DPDs that color interference terms is suppressed by Sudakov factors at large scales \cite{Mekhfi:1985dv,Manohar:2012jr}, which when combined with positivity bounds constraining the size of the correlations at low scales \cite{Kasemets:2014yna} set limits on the scale at which these correlations can be of experimental relevance. The scale evolution of the DPDs are described by generalizations of the usual DGLAP evolution equations. Two versions of this have been discussed in the literature: a homogeneous equation describing the separate evolution of each of the two partons and an inhomogeneous including also the splitting of one parent parton into the two partons that undergo hard scattering \cite{Kirschner:1979im,Shelest:1982dg,Snigirev:2003cq,Gaunt:2009re,Ceccopieri:2010kg}. Which
version is adequate for the description of DPS
processes remains controversial in the literature
\cite{Gaunt:2011xd,Gaunt:2012dd,Diehl:2011tt,Diehl:2011yj,%
  Ryskin:2011kk,Manohar:2012pe,Blok:2011bu,Blok:2013bpa}.

These proceedings reviews our study of the effects scale evolution has on DPS correlations \cite{Diehl:2014vaa}. For the numerical results presented in these proceedings we have used the homogeneous evolution equation. To solve the evolution equations numerically, we use a modified version of
the code originally described in \cite{Gaunt:2009re}.

\section{Correlations between $x_1, x_2$ and $y$}
\label{sec:trans}
A number of arguments suggest an
interplay between the dependence of DPDs on the longitudinal momentum
fractions $x_1,$ $x_2$ of the partons, as well as between their momentum
fractions and their relative transverse distance $\vek{y}$
\cite{Diehl:2013mma}. 

In this section we study the impact of evolution on the correlations between the momentum fractions and the interparton distance, for this purpose we need a model for the DPDs at the starting scale of evolution. We take a simple ansatz motivated by studies of GPDs as explained in  \cite{Diehl:2014vaa} 
\begin{align}
  \label{eq:g-ansatz}
F_{ab}(x_1,x_2,\y) =f_a(x_1) \ms f_b(x_2)\,
   \frac{1}{4\pi h_{ab}(x_1,x_2)}\,
      \exp\biggl[ - \frac{\y^2}{4\ms h_{ab}(x_1,x_2)} \biggr]\, ,
\end{align}
at the starting scale $Q_0^2 = 2 \gev^2$, with
\begin{align}
  \label{eq:dpd-trans}
h_{ab}(x_1,x_2) = h_a(x_1) + h_b(x_2)
 = \alpha_{a}'\ln\frac{1}{x_1} + \alpha_{b}'\ln\frac{1}{x_2}
   + B_{a} + B_{b} \,.
\end{align}
The parameters are set to the values
\begin{align}
  \label{eq:trans-param}
  \alpha_{q^-}' & = 0.9 \gev^{-2}\,, &
   \alpha_{q^+}' & = 0.164 \gev^{-2}\,, &
 \alpha_{g}'   & = 0.164 \gev^{-2}\,, 
\nonumber \\
 B_{q^-}       & = 0.59 \gev^{-2}\,, &
  B_{q^+}       & = 2.4 \gev^{-2}\,, &
    B_{g}         & = 1.2 \gev^{-2} \,.
\end{align}


The DPDs evolve independently at each
value of $y$, but the interplay
between $y$ and the momentum fractions $x_1$ and $x_2$ in the starting
conditions has consequences for the scale evolution at
different values of $y$. 

\begin{figure}[htb]
  \centering
	\includegraphics[width=0.465\textwidth]{%
      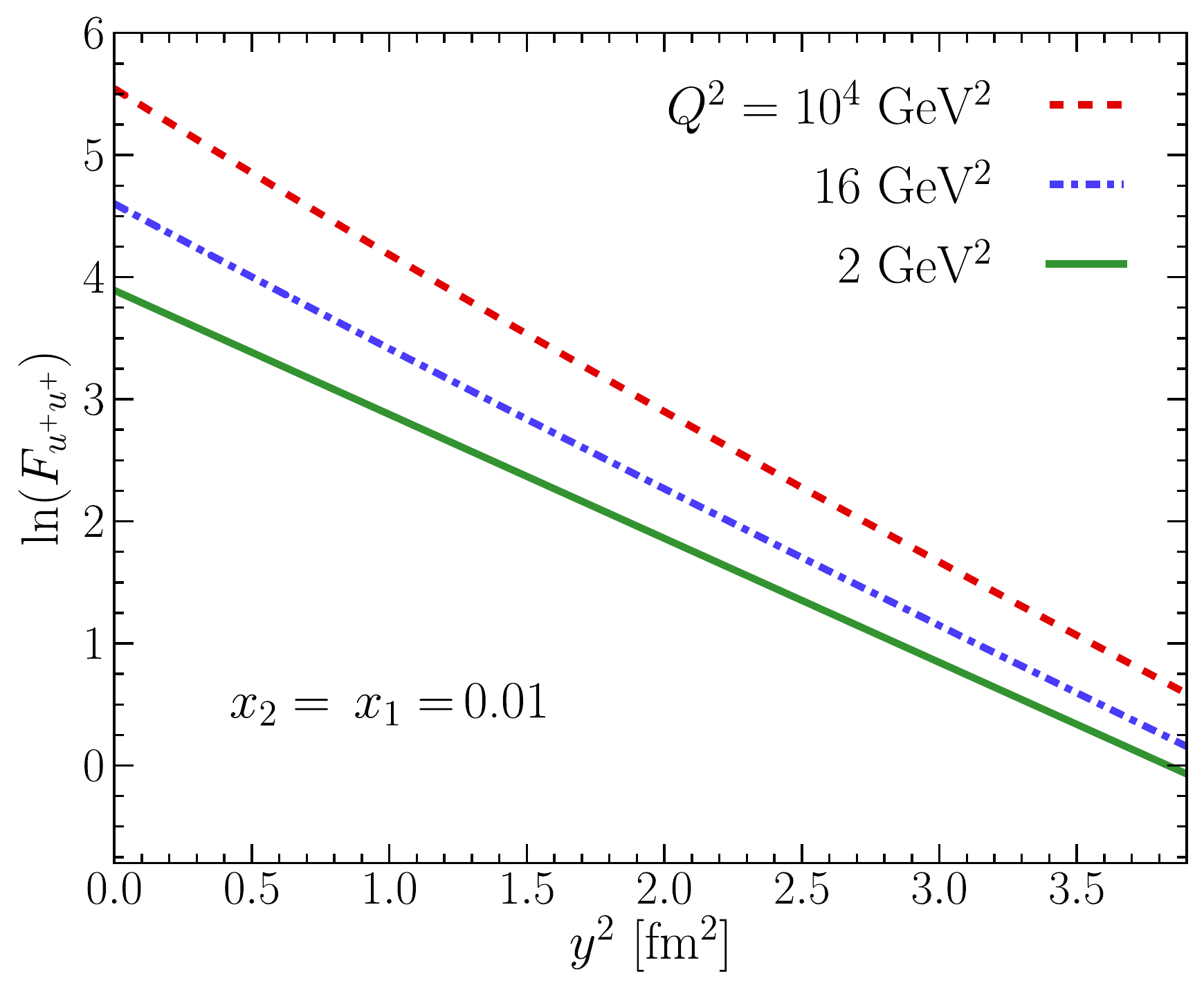}
 	\includegraphics[width=0.515\textwidth]{%
      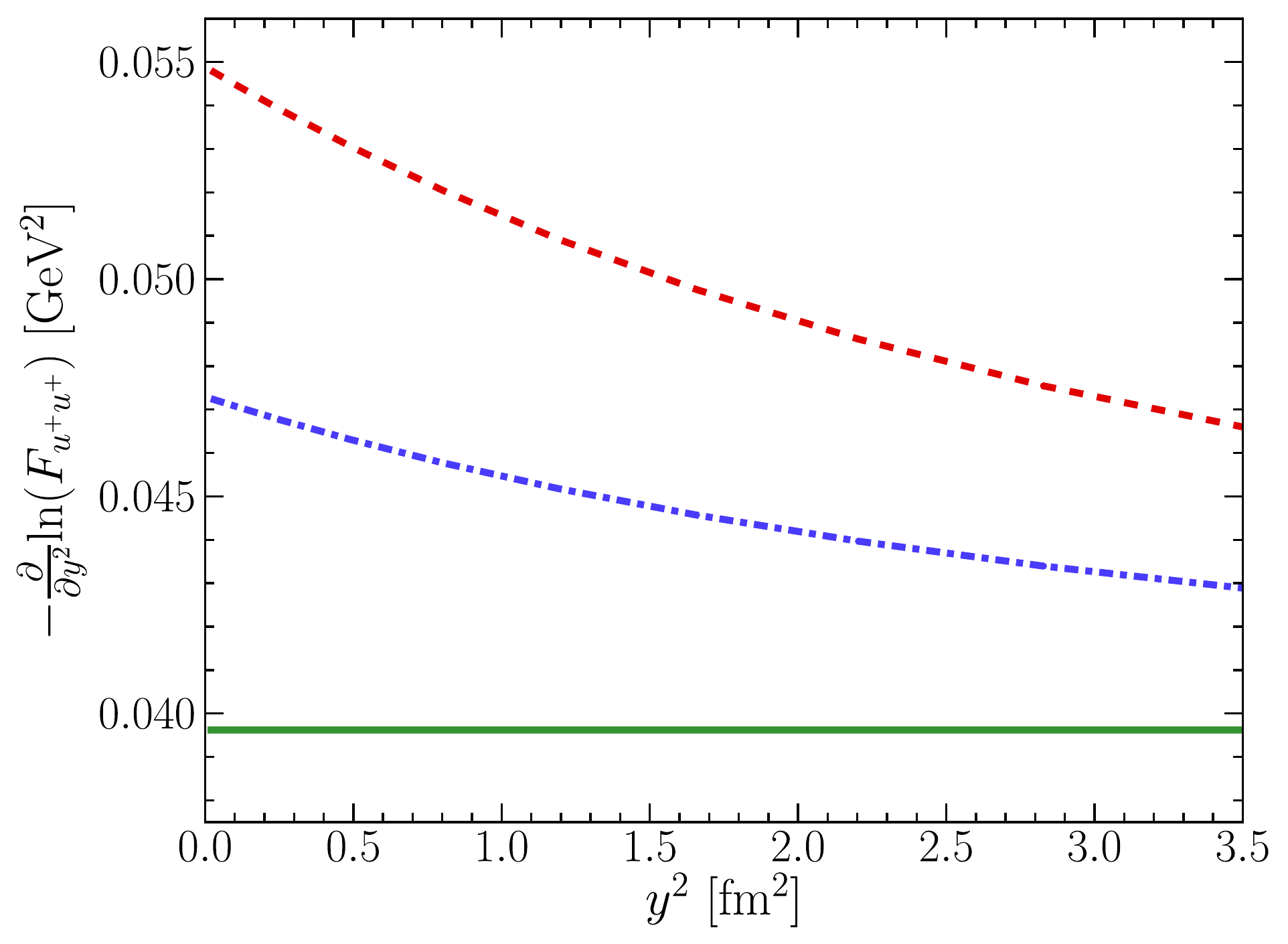}
\vspace*{8pt}
  \caption{\label{fig:ydep} $y^2$ dependence of the DPD for two $u^+$.  The left panel
    shows the natural logarithm of the DPD and the right panels the
    corresponding slope in $y^2$.  Longitudinal momentum fractions are
    fixed at $x_1 = x_2 = 0.01$.}
\end{figure}

\begin{figure}[tb]
  \centering
  \includegraphics[width=0.49\textwidth]{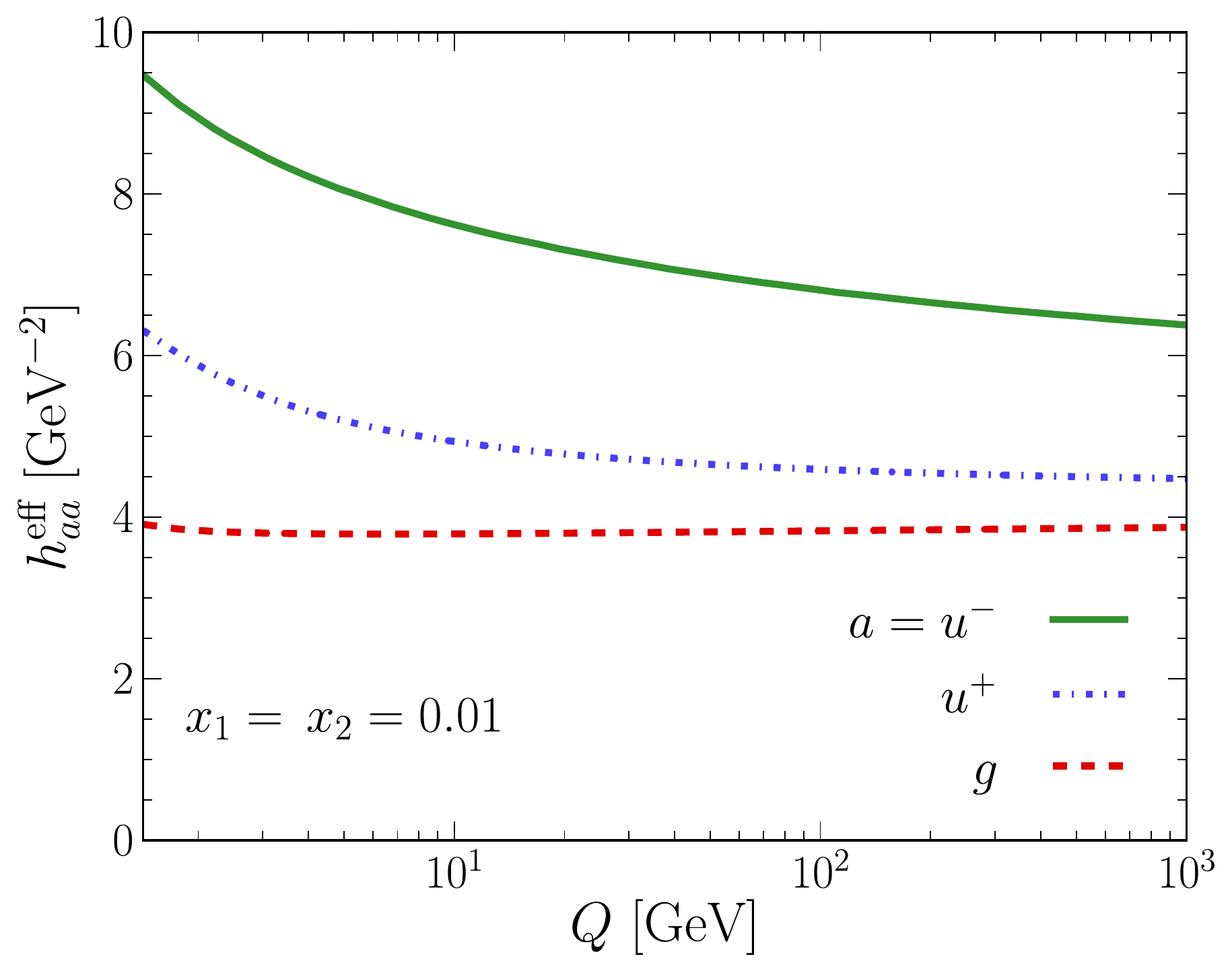}
  \includegraphics[width=0.493\textwidth]{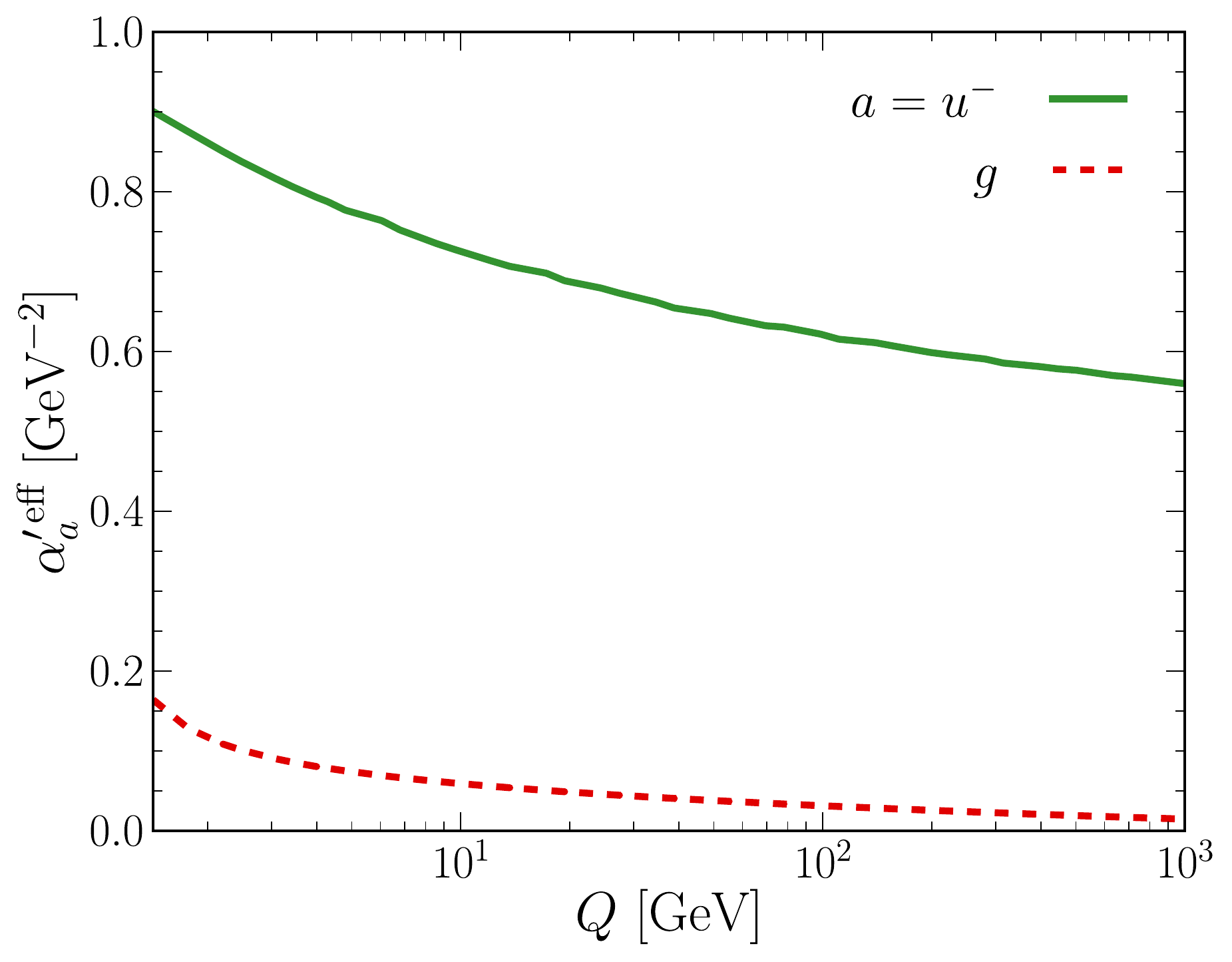}
\vspace*{8pt}
  \caption{\label{fig:fit-result-params} (a): Evolution of the effective
    Gaussian width $h_{aa}^{\text{eff}}(x,x)$ and evaluated at $x = 0.01$ for $a=u^+$,
    $u^-$ and $g$.  (b): Evolution of the effective shrinkage parameter
    $\alpha'^{\,\text{eff}}_a$ obtained by fitting
    $h_{aa}^{\text{eff}}(x,x)$ in the range $0.004
    \le x \le 0.04$ for $a = u^-$ and $g$.}
\end{figure}

The Gaussian dependence of our starting condition \eqref{eq:g-ansatz} is approximately preserved by evolution up to large scales, as demonstrated by figure~\ref{fig:ydep}. This allows us to take a closer look at the evolution of the width of the $y$
dependence. Figure~\ref{fig:fit-result-params}(a) shows the evolution of
$h_{aa}^{\text{eff}}(x,x)$ (effective Gaussian width) at $x = 0.01$ for $a=u^-$, $u^+$ and $g$.  
The effective
Gaussian width decreases under evolution for both $u^-$ and $u^+$, whereas
it changes for the gluon. As the valence combination $u^-$ evolves to
higher scales, partons move from higher to lower $x$ values by
radiating gluons.  For partons at given $x$ and $Q$, the width of the $y$
distribution is therefore influenced by the smaller values of this width for
partons with higher $x$ at lower $Q$ - leading to the decrease of
$h_{u^-u^-}^{\text{eff}}$ with $Q$ in
figure~\ref{fig:fit-result-params}(a).
The double $u^+$ distribution mixes with gluons and $h_{u^+u^+}^{\text{eff}}$ approaches $h_{gg}^{\text{eff}}$
with increasing scale, although it does so rather slowly.  The difference
between the transverse distribution of gluons and quarks, which we have
assumed at $Q_0$, remains up to large scales.

\begin{figure}[tb]
  \centering
  \includegraphics[width=0.28\textwidth]{%
      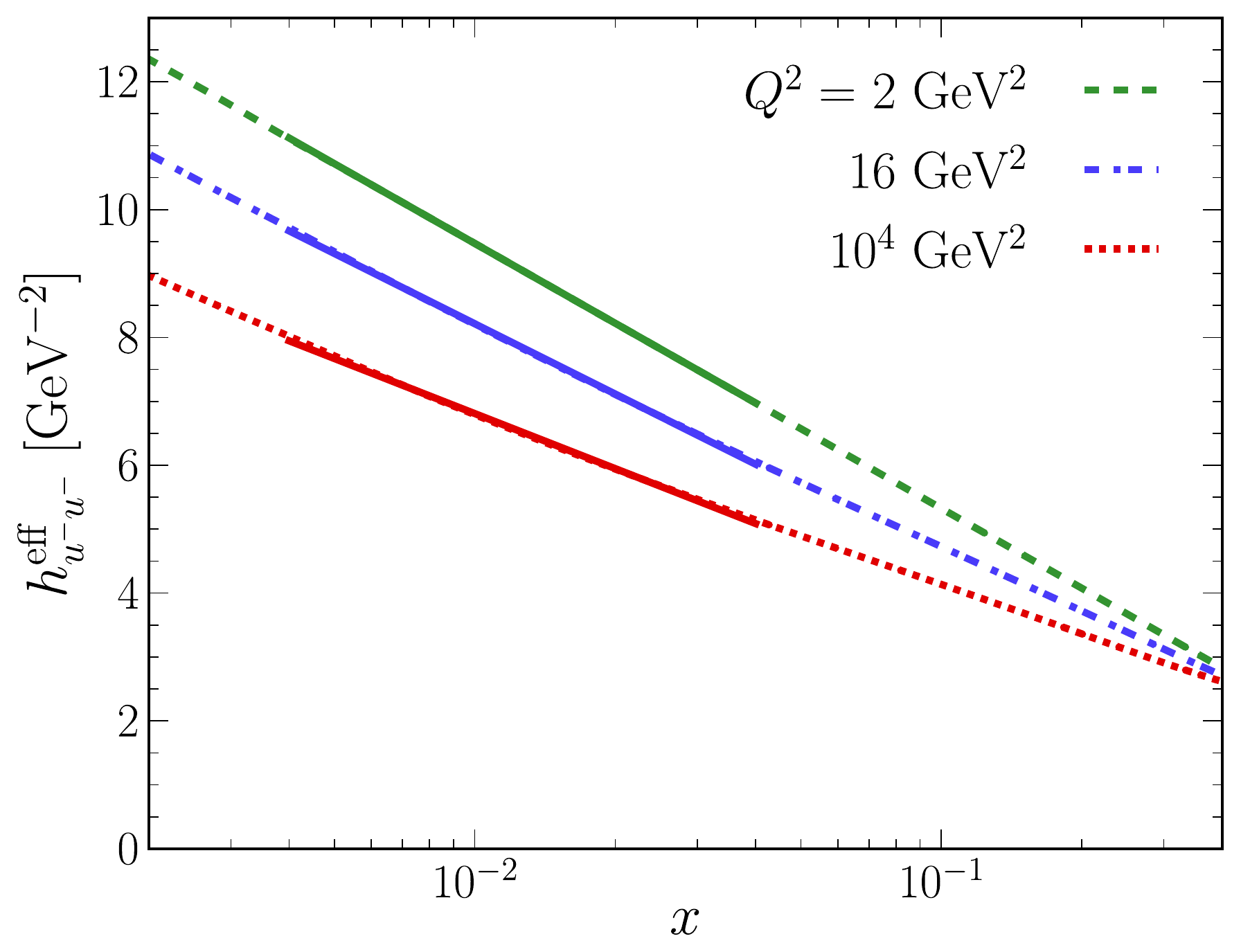}
  \includegraphics[width=0.28\textwidth]{%
      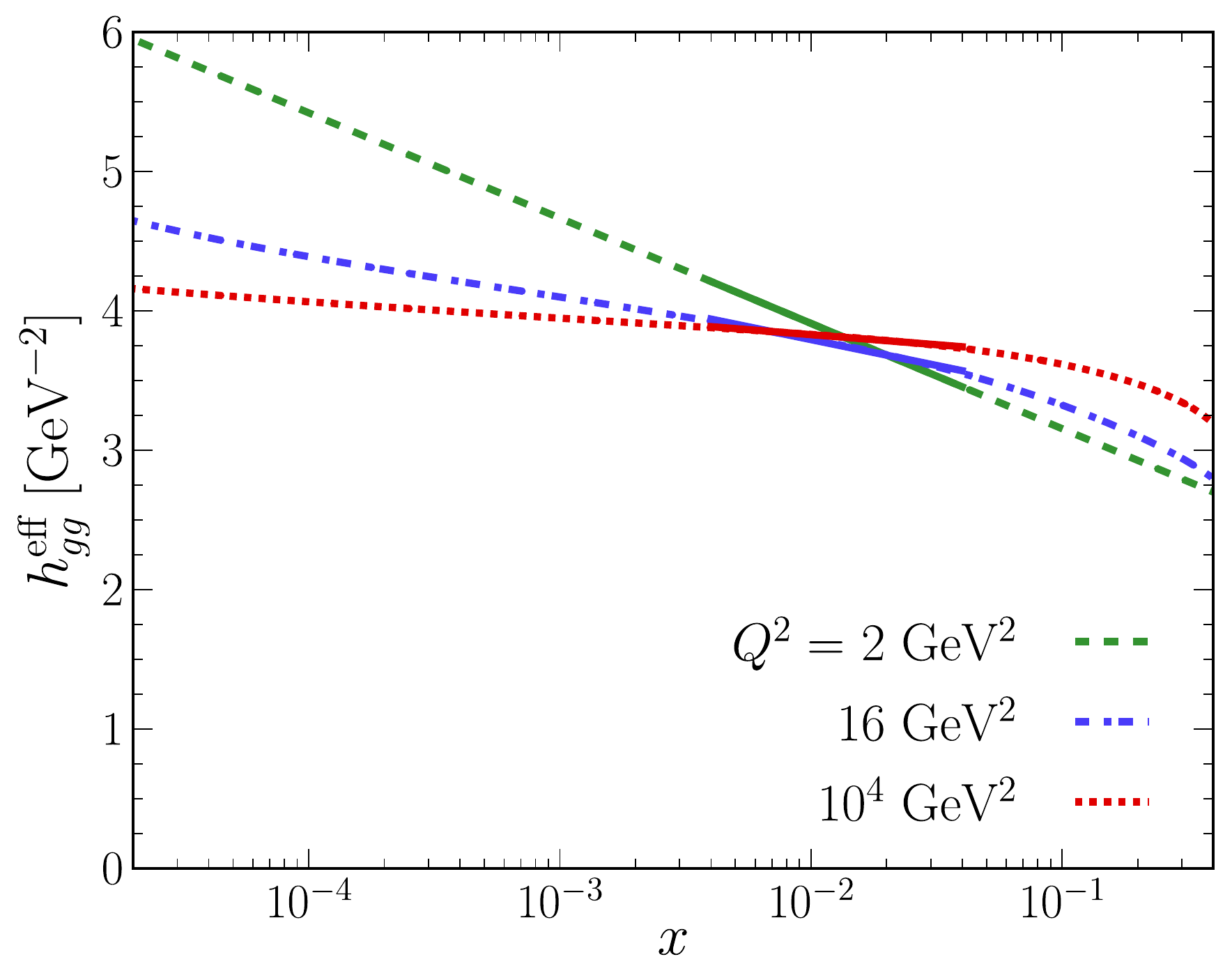}
  \includegraphics[width=0.28\textwidth]{%
      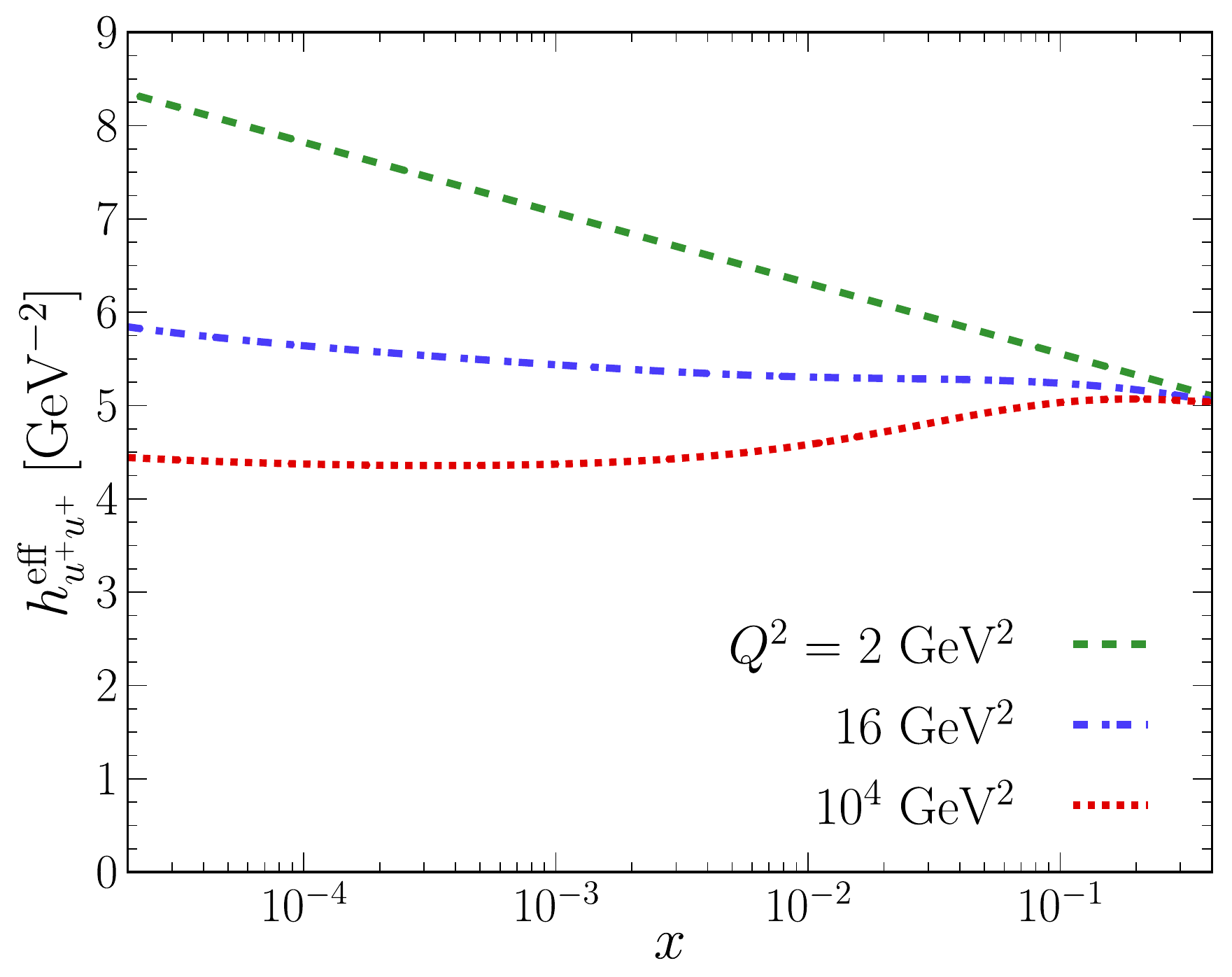}
\vspace*{8pt}
  \caption{\label{fig:fit-test}: Dependence of
    $h_{aa}^{\text{eff}}(x,x)$ on $x$.  The solid sections of the curves
    in panels (a) and (b) represent fits in figure~\ref{fig:fit-result-params} in
    the range $0.004 \le x \le 0.04$. }
\end{figure}

The dependence of $h_{aa}^{\text{eff}}(x,x)$ on $x$ is shown in
figure~\ref{fig:fit-test}(a), (b) and (c) for the different parton types.
We see that the evolution is faster at smaller momentum fractions $x$, and at low $x$ there is a rapid decrease of
$h_{aa}^{\text{eff}}(x,x)$ with $Q^2$ for all parton types.  For $u^+$
this results in a region of intermediate $x$ where
$h_{u^+u^+}^{\text{eff}}(x,x)$ increases with $x$ at high $Q^2$.  For
$u^-$ and $g$ the curves for $h_{aa}^{\text{eff}}(x,x)$ are approximately
linear in $\ln(x)$ as long as we stay away from the large-$x$ region.
This allows us to extract an effective shrinkage parameter
$\alpha'^{\,\text{eff}}_a$ by fitting the effective Gaussian width
in an appropriate region of $x$. The scale dependence of
$\alpha'^{\,\text{eff}}_a$ is shown in
figure~\ref{fig:fit-result-params}(b).  We find that
$\alpha'^{\,\text{eff}}_a$ decreases quite rapidly for $a=g$ and more
gently for $a=u^-$.

In summary, we find that a
Gaussian $y$ dependence at the initial scale is approximately preserved
under evolution, with a noticeable but relatively slow change of the
effective Gaussian width.  Despite the mixing between gluons and quarks in
the singlet sector, the differences between their distributions remains up
to high scales.

\section{Evolution of Polarized Double Parton Distributions}
\label{sec:pol}

We next investigate the evolution of spin correlations between
two partons inside a proton.  For this purpose we assume  a
multiplicative $y$ dependence of the DPDs, 
\begin{align}
  \label{eq:y-factorization}
f_{p_1 p_2}(x_1,x_2,\y; Q) &=
   \tilde{f}_{p_1 p_2}(x_1,x_2; Q) \,G(\y) \,,
\end{align}
As our focus is on the degree
of parton polarization rather than on the absolute size of the DPDs, we
set $G(\y)=1$. For the unpolarized DPDs we take a simple factorizing
ansatz at the starting scale ($Q_0^2 = 1 \gev^2$),
\begin{align}
  \label{eq:x-factorization}
\tilde{f}_{ab}(x_1,x_2; Q_0) &= f_a(x_1; Q_0)\, f_b(x_2; Q_0) \,.
\end{align}
The single parton densities used will be either of the two
LO sets MSTW 2008 \cite{Martin:2009iq} and GJR 08 \cite{Gluck:2007ck}.

To model the polarized DPDs is more difficult. There is no reason to believe that a decomposition of a polarized DPD, which describes the spin correlations between two partons, into polarized PDFs, describing the spin correlations between the proton and a parton, should be suitable even as a starting point. Instead in the scenario presented in these proceedings, we make use of the positivity bounds for DPDs derived in
\cite{Diehl:2013mla}.  At the starting scale $Q_0$ of evolution, we
maximize each polarized DPD with respect to its unpolarized
counterpart. This gives polarized distributions equal to the unpolarized at the initial scale.

We will show a series of figures with curves for different scales.  In each figure, the upper row shows the
polarized DPDs and the lower row shows the ratio between polarized and
unpolarized DPDs.  The
ratio indicates how important spin correlations are in the cross
sections of DPS processes.  We show the polarized
distributions as functions of $x_1$ at $x_1=x_2$ and as functions of
$\ln(x_1/x_2)$ for $x_1x_2=10^{-4}$. 

\subsection{Quark distributions}
We start our examination of spin correlations with the DPDs for
longitudinally or transversely polarized quarks and antiquarks. 

\begin{figure}[tb]
  \centering
   	\includegraphics[width=0.49\textwidth]{%
      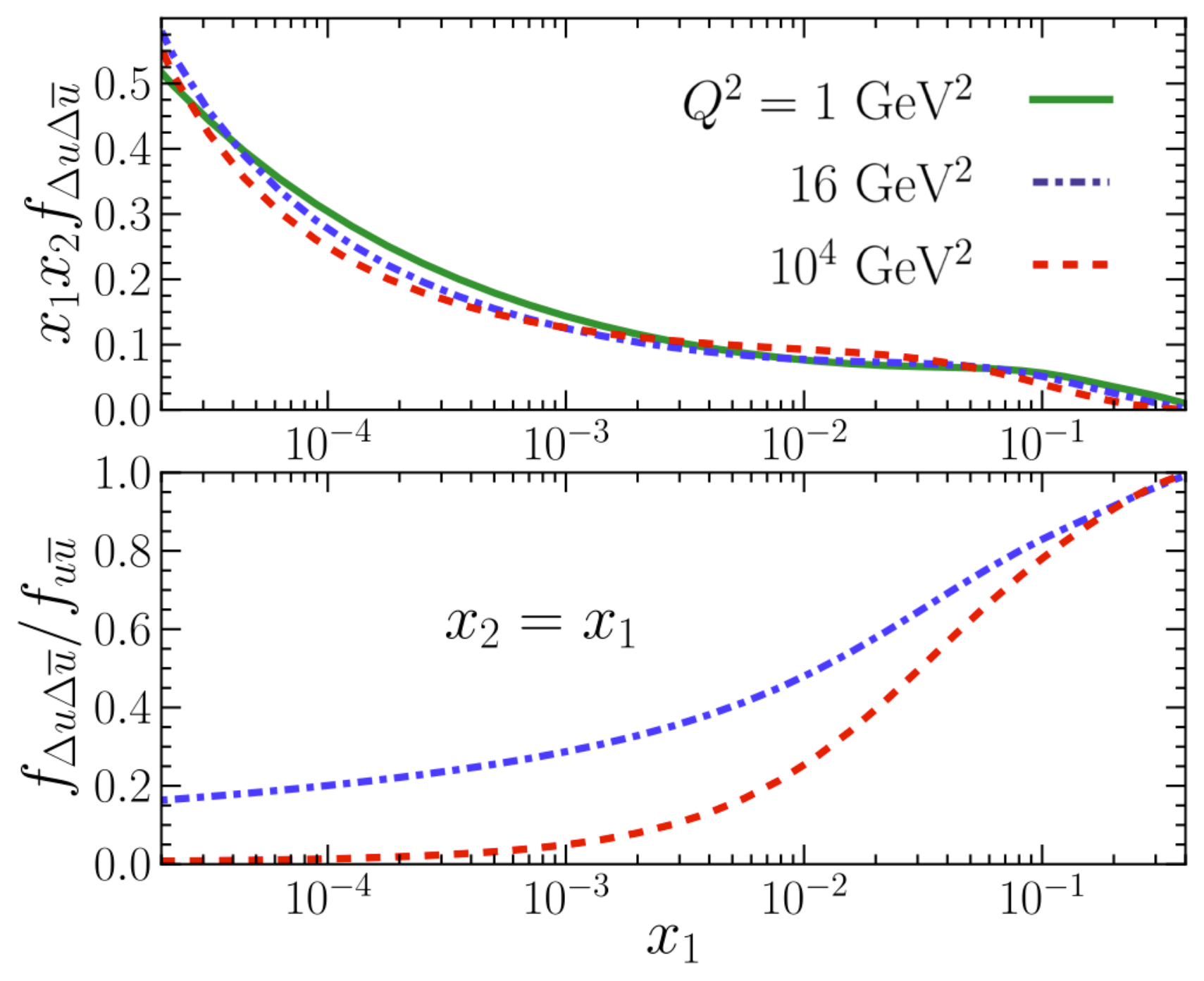}
	\includegraphics[width=0.49\textwidth]{%
      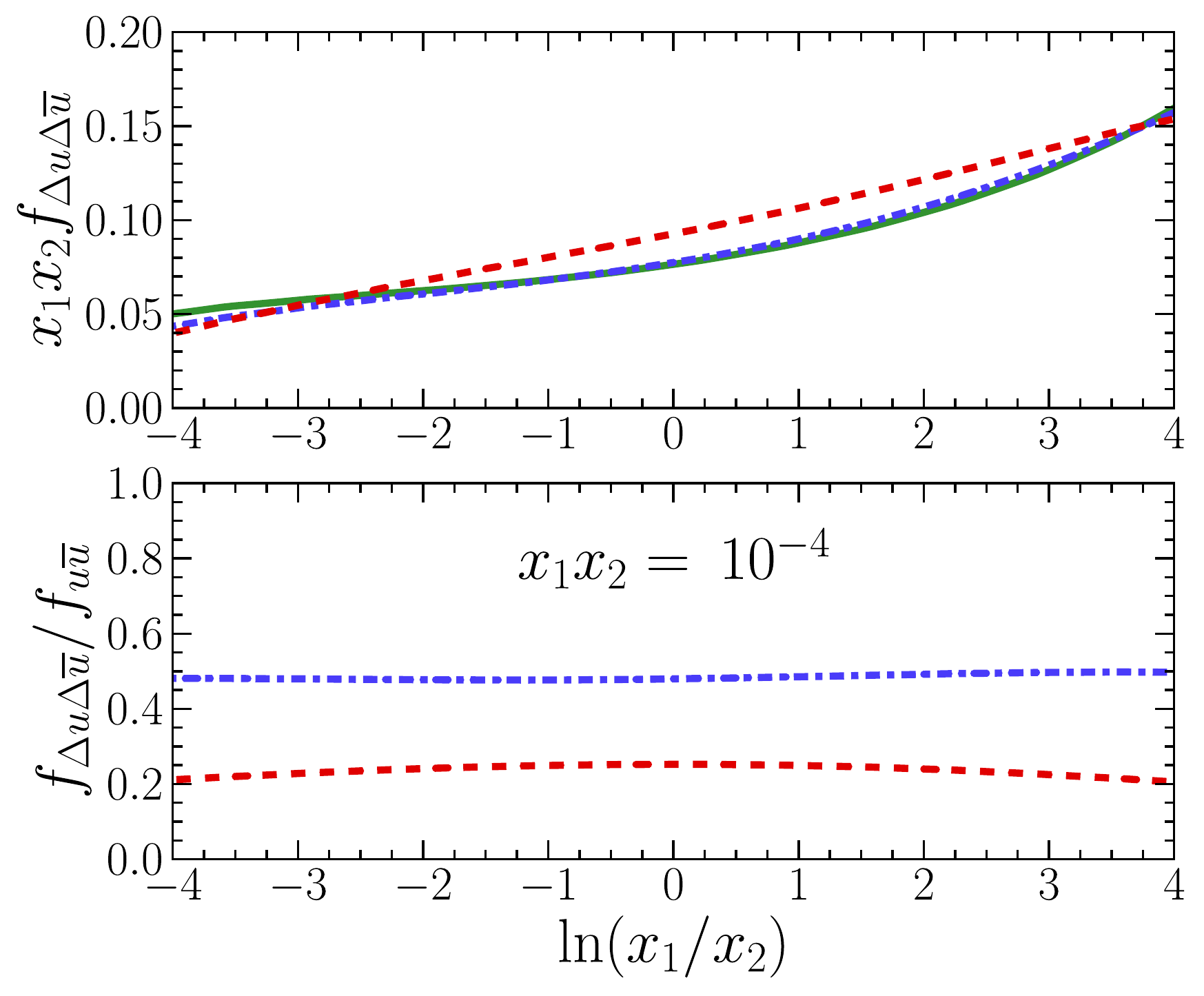}
      \vspace*{8pt}
  \caption{\label{fig:long-u-max} Longitudinally polarized up quarks and
    antiquarks, with initial conditions using the MSTW
    PDFs. Here and in the following figures the upper row shows the
    polarized DPDs and the lower row the ratio between polarized and
    unpolarized DPDs. }
\end{figure}

The distribution for longitudinally polarized up quarks and antiquarks is shown in figure~\ref{fig:long-u-max}.  The polarized
distribution $f_{\Delta u\Delta\bar{u}}$ evolves very slowly, but the degree of polarization decreases with
the evolution scale.  This is due to the increase of the unpolarized DPDs. We find a
degree of polarization around 50\% at $Q^2=16 \gev^2$ and above 20\% at
$Q^2=10^4 \gev^2$ for $x_1\ms x_2=10^{-4}$ and a wide range of
$\ln(x_1/x_2)$.

\begin{figure}[tb]
  \centering
  \includegraphics[width=0.49\textwidth]{%
      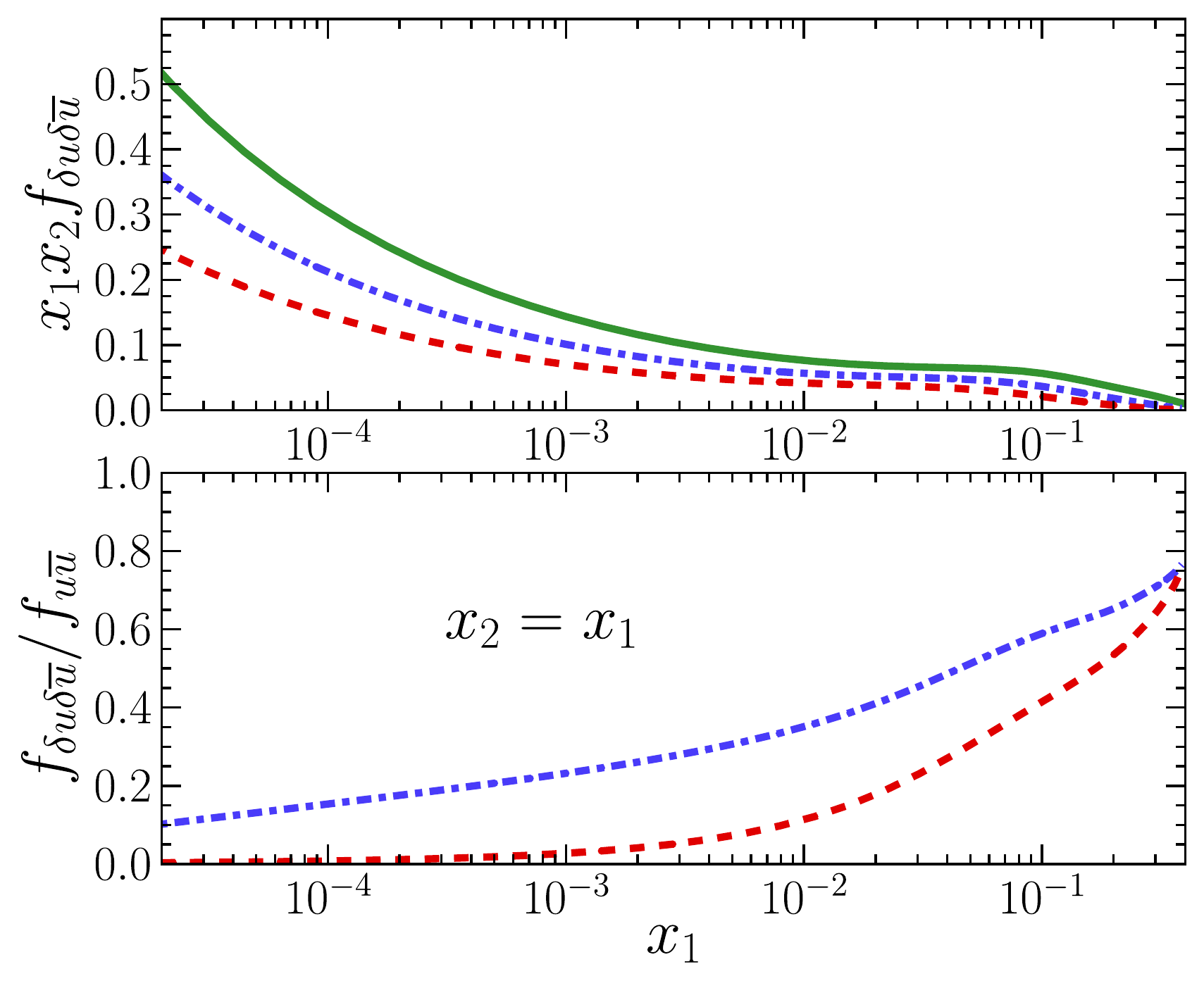}
  \includegraphics[width=0.49\textwidth]{%
      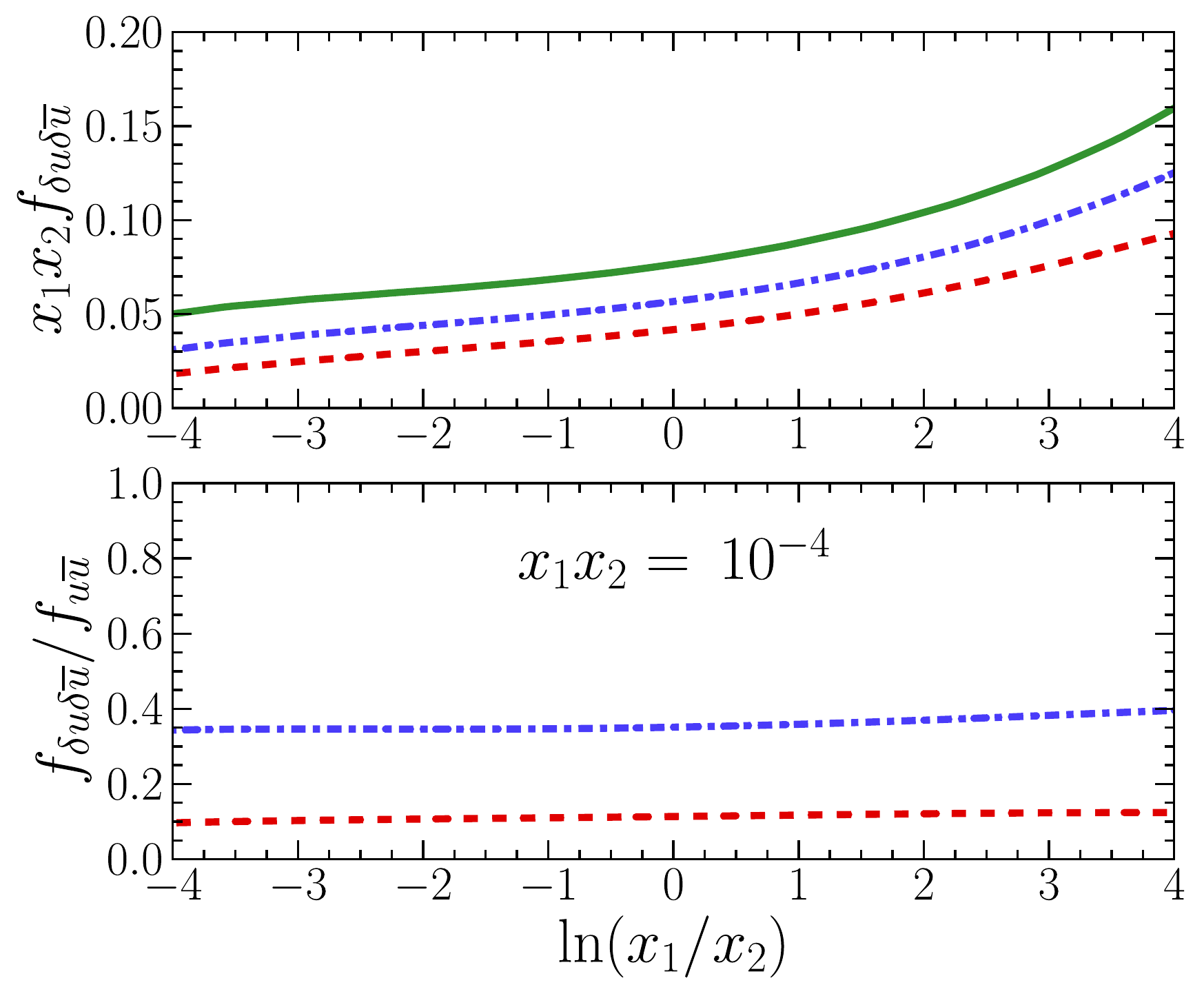}
      \vspace*{8pt}
  \caption{\label{fig:trans-u-max} Transversely polarized up quarks and
    antiquarks, with initial conditions using the MSTW
    PDFs. Color (line style) coding as in figure~\ref{fig:long-u-max}.}
\end{figure}

Transverse quark and antiquark polarization leads to
characteristic azimuthal correlations in the final state of DPD processes
\cite{Kasemets:2012pr}.  They do
not mix with gluons under evolution, nor with quarks or antiquarks of
different flavors.  Figure~\ref{fig:trans-u-max} shows the DPD for
transversely polarized up quarks and antiquarks.
There is a small decrease of the DPD with $Q^2$ over the entire $x_i$
range, but the suppression of the degree of polarization is mainly due to
the increase in the unpolarized distributions.  The evolution of the
degree of polarization is similar to the case of longitudinal polarization, with a somewhat faster decrease.  At intermediate and
large $x_i$ values, the degree of polarization decreases slowly.  For
$x_1x_2=10^{-4}$ it amounts to 40\% at $Q^2=16 \gev^2$ and to 10\% at
$Q^2=10^4 \gev^2$ over a wide rapidity range.

The polarization for other combinations of light quarks and antiquarks is
of similar size and shows a similar evolution behavior as for the case of
a $u\bar{u}$ pair. 


\begin{figure}[b]
  \centering
  \includegraphics[width=0.496\textwidth]{%
      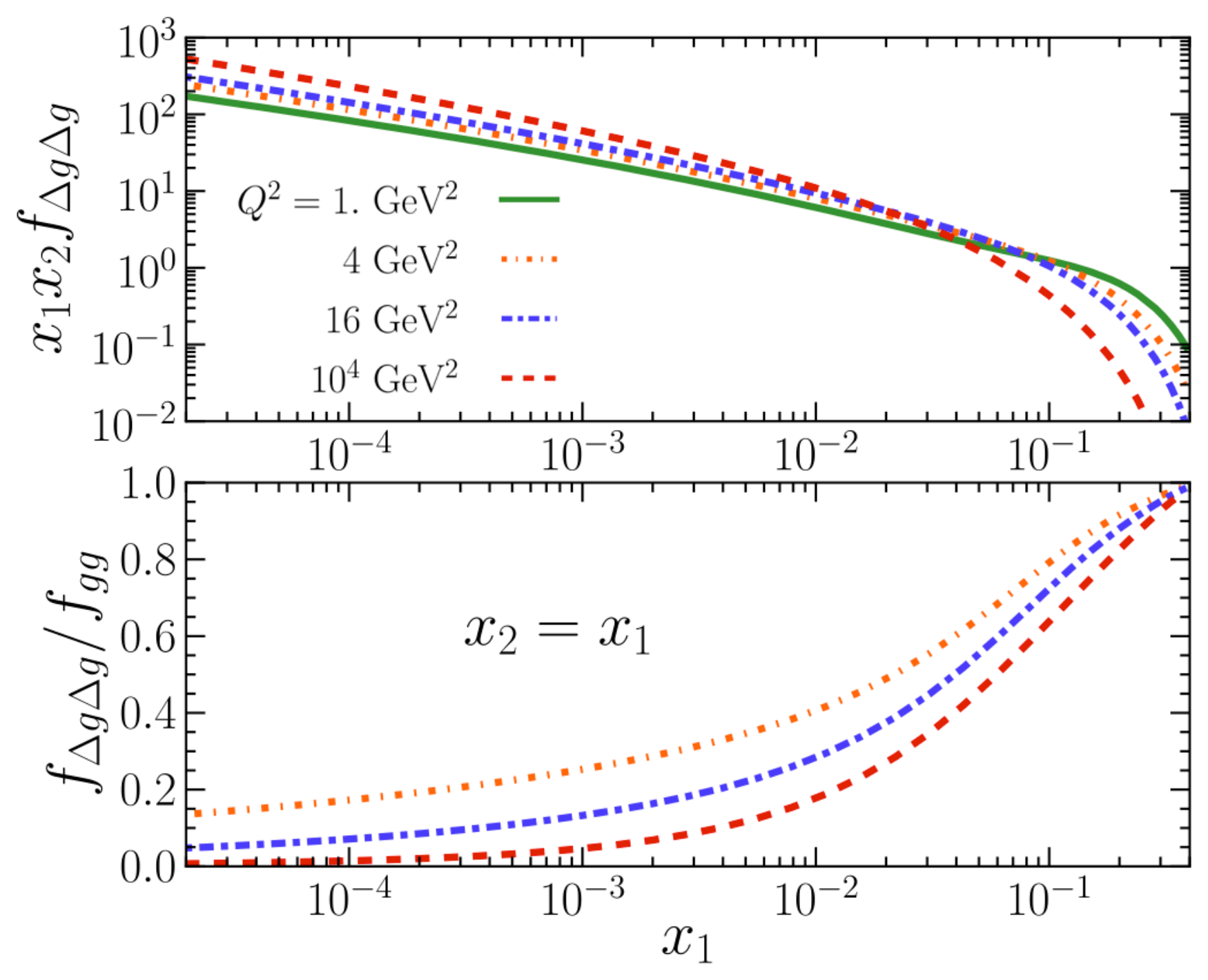}
  \includegraphics[width=0.48\textwidth]{%
      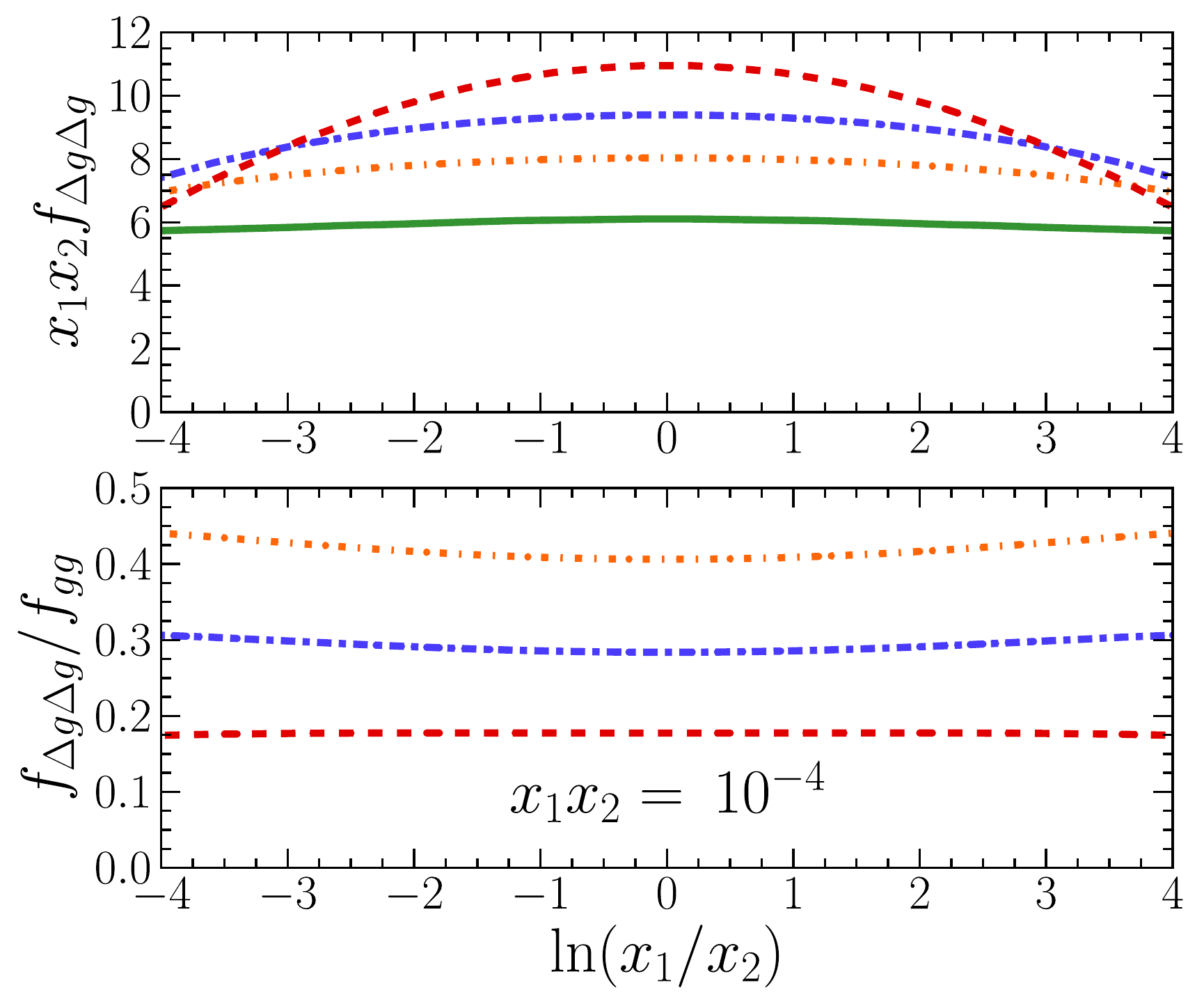}
\vspace*{8pt}
  \caption{\label{fig:long-gGJR-max} Distribution for two longitudinally polarized gluons with initial conditions using the PDFs of GJR.}
\end{figure}

\subsection{Gluon distributions}
Gluons can be polarized longitudinally or linearly.  The unpolarized (single or double) gluon density increases
rapidly at small momentum fractions due to the $1/x$ behavior of the gluon
splitting kernel.  The absence of this low-$x$ enhancement in the
polarized gluon splitting kernels lead us to expect that the degree of
gluon polarization will vanish rapidly in the small $x$ region. 

 As can be
seen in figure~\ref{fig:long-gGJR-max} for longitudinally polarized
gluons, this is indeed the case.  The distribution $f_{\Delta g \Delta g}$
does increase with evolution scale, but at a much lower rate than
$f_{gg}$.  Evolution quickly suppresses the degree of longitudinal
gluon polarization in the small $x_i$ region. The degree of polarization at at $Q^2=16 \gev^2$ amounts to a degree of polarization equal to 30\% 
 and almost 20\% at $Q^2=10^4 \gev^2$ for $x_1\ms
x_2=10^{-4}$, with a very weak dependence on $\ln(x_1/x_2)$. Our knowledge of the single gluon distribution at the low scale remains poor, and using as an alternative to the GJR distributions the MSTW set the degree of polarization is reduced to around half the size or below.

\begin{figure}[tb]
\centerline{
\includegraphics[width=0.496\textwidth]{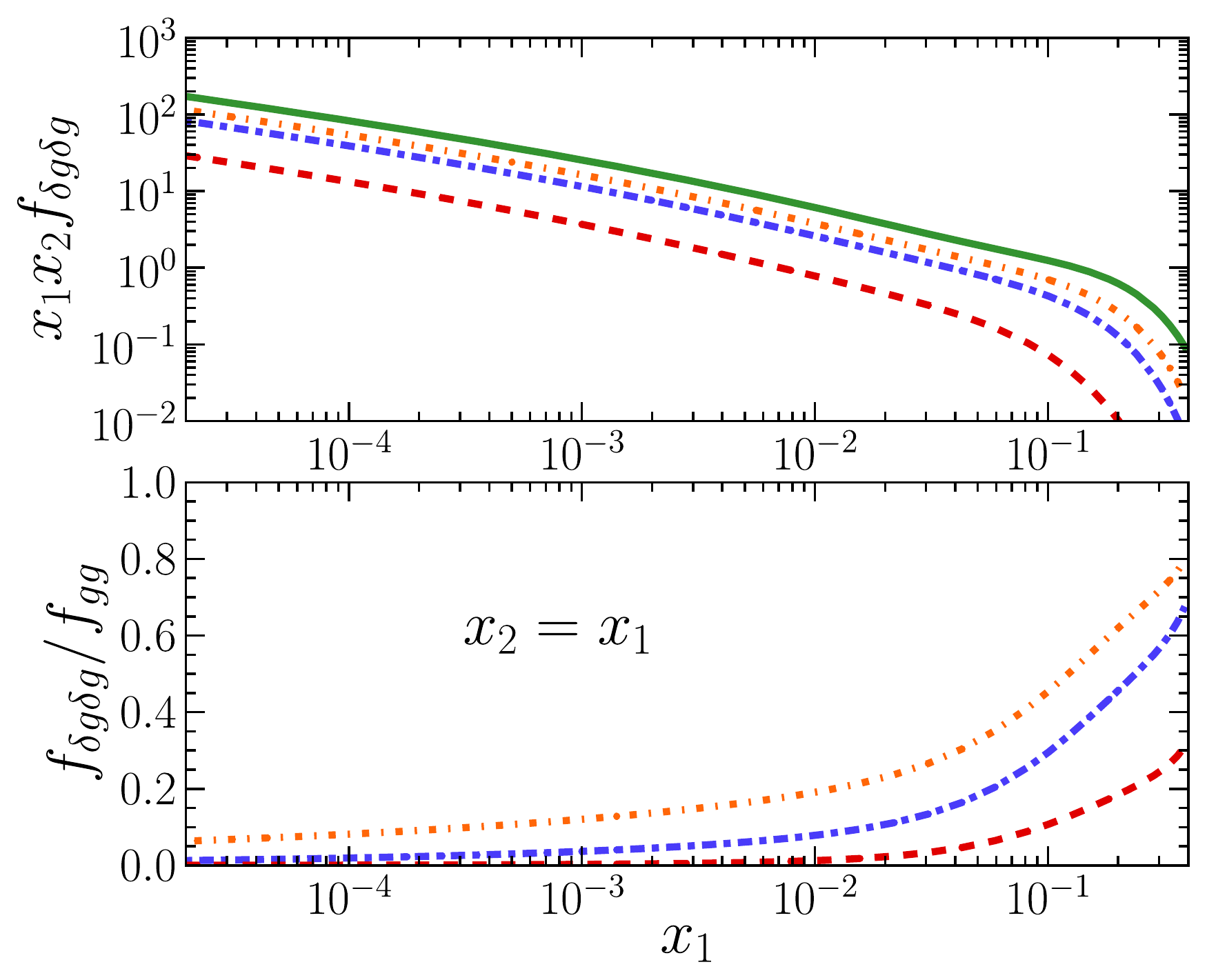}
\includegraphics[width=0.48\textwidth]{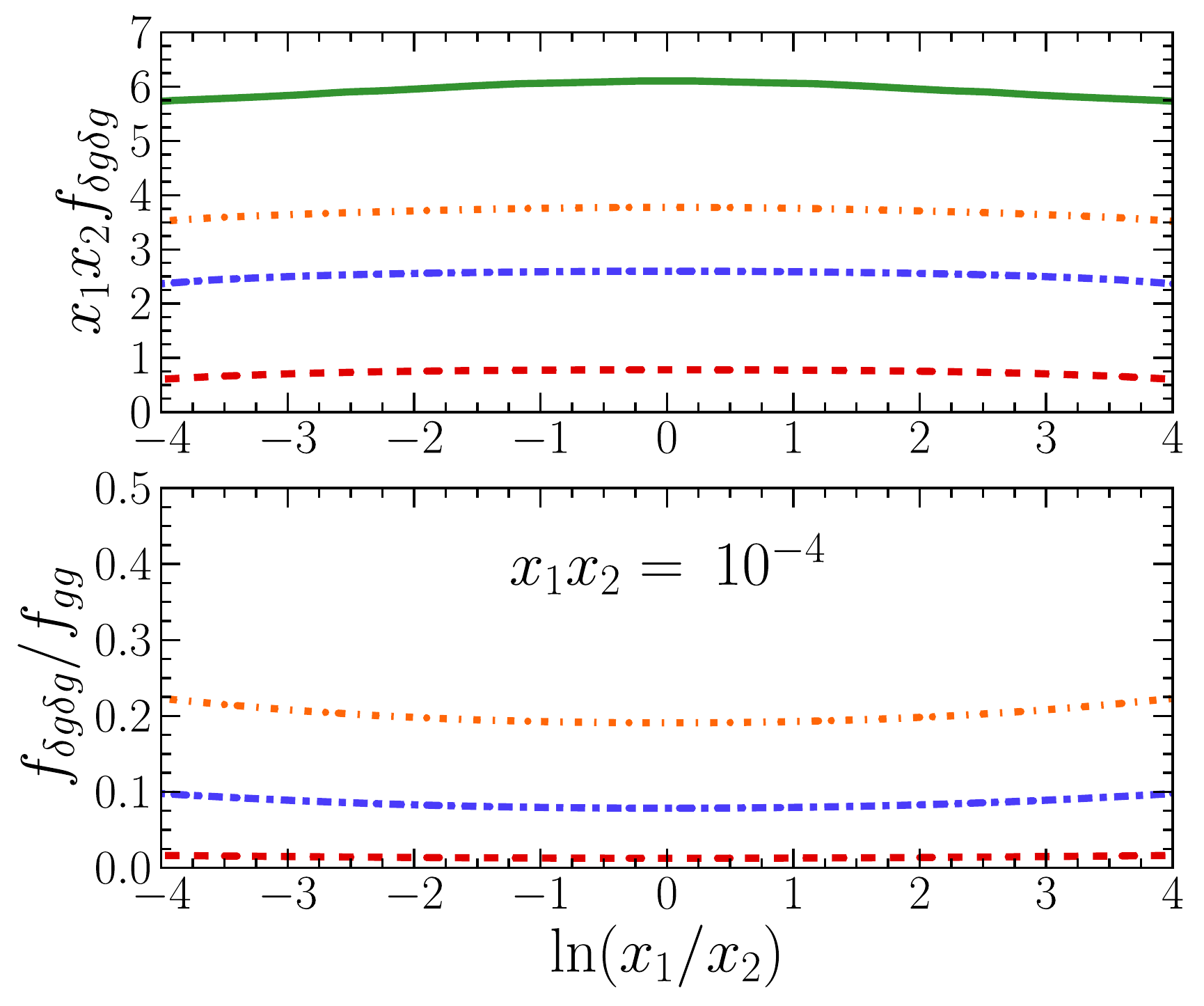}
}
\vspace*{8pt}
\caption{\label{fig:lin-g-max} Distribution for two linearly polarized
    gluons using the GJR PDFs in the initial
    conditions. Color (line style) coding as in
    figure~\protect\ref{fig:long-gGJR-max}.}
\end{figure}

Linearly polarized gluons give rise to azimuthal asymmetries in DPS cross
sections \cite{Kasemets:2014xxx}.  The effect of evolution on the distribution of two linearly
polarized gluons is shown in
figure~\ref{fig:lin-g-max}.  We see that even the polarized distribution
$f_{\delta g \delta g}$ itself decreases with the scale.  Together with
the rapid increase of the unpolarized two-gluon DPD this results in a
rapid decrease of the degree of linear polarization, especially at small
$x_i$.  As in the case of longitudinal gluon polarization, using the MSTW
distributions at the starting scale results in an even
faster suppression.  In that case the degree of polarization is tiny
already at $Q^2=16 \gev^2$.  We conclude that the correlation between two linearly polarized
gluons is quickly washed out by evolution and can only be relevant at
rather large $x_i$ or rather low scales.

\providecommand{\href}[2]{#2}\begingroup\raggedright\endgroup
\end{document}